\documentclass[prl,twocolumn,showpacs]{revtex4}
\usepackage[dvips]{graphicx}
\usepackage{latexsym}
\usepackage{amsmath}
\usepackage{amsfonts}
\usepackage{amssymb}
\usepackage{bm}
\begin{document}
\newcommand{\fig}[2]{\includegraphics[width=#1]{#2}}
\newcommand{\la}{\langle}
\newcommand{\ra}{\rangle}
\newcommand{\dg}{\dagger}
\newcommand{\upa}{\uparrow}
\newcommand{\dna}{\downarrow}
\newcommand{\as}{{\alpha\sigma}}
\newcommand{\hH}{{\hat{\mathcal{H}}}}
\newcommand{\hn}{{\hat{n}}}
\newcommand{\hP}{{\hat{P}}}
\newcommand{\bk}{{{\bf k}}}
\newcommand{\bq}{{{\bf q}}}
\newcommand{\bQ}{{{\bf Q}}}
\setlength{\unitlength}{1mm}

\title{Electron correlation and spin density wave order in iron pnictides}

\author{Sen Zhou$^1$ and Ziqiang Wang$^2$}
\affiliation{$^1$ National High Magnetic Field Laboratory, Florida
State University, Tallahassee, Florida 32310, USA} \affiliation{$^2$
Department of Physics, Boston College, Chestnut Hill, Massachusetts
02467, USA}

\date{\today}

\begin{abstract}

We study the correlation effects on the electronic structure and
spin density wave order in Fe-pnictides. Using the multiorbital
Hubbard model and Gutzwiller projection, we show that
correlation effects are essential to stabilize the
metallic spin density wave phase for the intermediate correlation
strengths appropriate for pnictides. We find that the ordered
moments depend sensitively on the Hund's rule coupling $J$ but
weakly on the intraorbital Coulomb repulsion $U$, varying from
$0.3\mu_B$ to $1.5\mu_B$ in the range $J=0.3\sim0.8$ eV for
$U=3\sim4$ eV. We obtain the phase diagram and discuss the effects
of orbital order and electron doping, the evolution of the Fermi
surface topology with the ordered moment, and compare to recent
experiments.

\typeout{polish abstract}
\end{abstract}
\pacs{71.27.+a, 74.70.Xa, 74.25.Ha, 74.25.Jb}

\maketitle

The iron pnictides have emerged recently as another class of
high-T$_c$ superconductors \cite{superconductivity} involving the
transition metal $d$-electrons in addition to the cuprates. In the
two most studied pnictide families, the 1111 (e.g.
LaO$_{1-x}$F$_x$FeAs) and the 122 series (e.g.
Ba$_{1-x}$K$_x$Fe$_2$As$_2$), the iron valence is Fe$^{2+}$. There
are six electrons occupying five Fe $3d$ orbitals. Their direct
overlap and via As $4p$ orbitals produce five energy bands with a
total bandwidth around 4eV \cite{singh08,kuroki08,Uvalue}. This is
comparable to the on-site Coulomb repulsion $U=3\sim4$eV
\cite{Uvalue} typical for transition metals. Thus, the Fe-pnictides
are multiorbital systems where the correlation strength is
intermediate and comparable to the kinetic energy. In this
intermediate correlation regime lies the challenge of the many-body
physics responsible for unconventional electronic ground states and
emergent phenomena in condensed matter and complex materials.

That the correlation effects play an important role in pnictides can
be seen from the fact that despite of the orbital degeneracy, the
normal state behaves quite incoherently with an enhanced magnetic
susceptibility in contrast to conventional Fermi liquids
\cite{haulekotliar}. At low temperatures, the observed quasiparticle
dispersion
\cite{ding08,malaeb08} exhibits a strong bandwidth reduction due to
electron correlations, as shown in a first principle calculation
that includes the interaction effects in the Gutzwiller approach
\cite{zfang09}. Appropriate treatment of the electron correlation in
this intermediate regime, especially its multiorbital nature, is
essential for understanding the properties, including the high-T$_c$
superconductivity, of these materials.

In this paper, we investigate the correlation effects on the spin
density wave (SDW) order from which the high-$T_c$ superconductivity
emerges with sufficient doping. At low temperatures, semi-metallic
SDW order develops in the undoped pnictides with an ordering vector
$\bQ=(\pi, 0)$ connecting the geometric centers of the electron and
hole Fermi surface (FS) pockets in the unfolded Brillouin zone (BZ)
containing one Fe atom per unit cell \cite{neutron1111, neutron122}.
The atomic states for Fe$^{2+}$ is predominantly $S=2$ and $S=1$ in
the presence of Hund's rule coupling and crystal field splitting.
However, the ordered Fe moments are much smaller than in the local
spin density approximation (LSDA)\cite{lsda} and vary substantially
from mostly about $0.35\mu_B$ in the 1111 series \cite{neutron1111}
to around $1\mu_B$ in the 122 series \cite{neutron122}. This
unconventional SDW phase evades the well established theories and
appears to straddle the limits of localized and itinerant magnetism.

We show that these unusual properties arise in the immediate
correlation regime, provided that the correlation effects are
treated appropriately. To this end, we study a multiorbital
Hubbard model for the Fe $t_{2g}$ complex with the hopping
parameters determined from the LDA band structure. The intra and
interorbital Coulomb repulsion $U$ and $U'$, and the Hund's rule
coupling $J$ are treated by Gutzwiller projection of multi-occupancy
in the intermediate correlation regime appropriate for the
pnictides. We show that the interplay between correlation and
itineracy stabilizes the metallic SDW phase. We found that $U$ and
$J$ play different roles in controlling the SDW order and the
bandwidth reduction. The ordered moments depend sensitively on
Hund's rule coupling but weakly on $U$ and varies from $0.3\mu_B$ to
$1.5\mu_B$ in the range $J=0.3\sim0.8$ eV and $U=3\sim4$ eV
consistent with experimental observations. The phase diagram is
obtained in the parameter space of $U$ and $J$ and contrasted to the
perturbative Hartree-Fock (HF) theory that erroneously predicts a
SDW metal-insulator transition (MIT). We elucidate the interplay
between the crystal field renormalization and the magnetization
induced band splitting and discuss the multiorbital nature of the
SDW and orbital order. The band dispersions and the FS topology are
shown to vary significantly as a function of the SDW moment with
implications for the ARPES experiments. We also present results for
the destruction of the SDW order with electron doping.

\begin{figure}
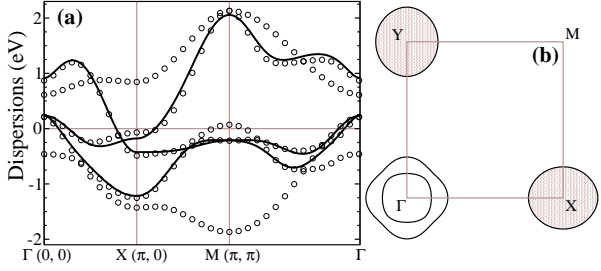

\begin{center}
\fig{3.1in}{fig1.eps} \vskip-3mm \caption{{\it Three-band tight-binding
model}. (a) Band dispersions (lines) and (b) Fermi surfaces. Open circles
in (a) denote the dispersions of the five-band model of Ref.
\cite{kuroki08}.} \label{fig1}
\end{center}
\vskip-8mm
\end{figure}

The Hamiltonian is written as $\hH=\hH_0+\hH_I$ where
\begin{equation}
\hH_0=\sum_{ij,\alpha\beta,\sigma}t_{\alpha\beta}[x_j-x_i,y_j-y_i]c^\dg_{i\as}
    c_{j\beta\sigma}+\sum_{i,\alpha}\Delta^0_\alpha \hn_{i\alpha}, \label{h0}
\end{equation}
describes the band structure of the Fe $3d$ complex that dominates
the low energy density of states in LDA \cite{Uvalue,kuroki08}. Here
$c^\dagger_{i\as}$ creates an electron in orbital $\alpha$ with spin
$\sigma$ on site $i$, $\hn_{i\alpha}$ is the density operator
and $\Delta^0_\alpha$ is the crystal field splitting. Since the
dispersions near the Fermi energy and the FS topology can be
captured by three $t_{2g}$ orbitals \cite{mazin08,leewen08}, we work
with a three-orbital model with $\alpha,\beta$=1 ($d_{xz}$), 2
($d_{yz}$), 3 ($d_{xy}$) derived from the five-band model of Kuroki
{\it et al.} for the LDA band dispersions of LaOFeAs
\cite{kuroki08}. The corresponding hopping integrals
$t_{\alpha\beta}[x_j-x_i,y_j-y_i]$ are given in Table \ref{hopping}
and $\Delta^0_\alpha=\{0, 0, \Delta_0=0.16\text{eV}\}$. The
dispersion and FSs are shown in Fig.~1 for the undoped case with
four electrons per Fe site. There are two hole pockets centered
around $\Gamma$ and two electron pockets around X and Y in the
unfolded BZ. Note that although the joint density of states is
enhanced, there is no FS nesting by the vector $\bQ$ and our results
are not sensitive to the details of the band parameters and the
shape of the FSs.

\begin{table}[htb]
\caption{\label{hopping} Hopping integrals $t_{\alpha\beta} [\Delta x,
\Delta y]$ up to five neighbors in unit of meV. Notations are the same as
in Ref. \cite{kuroki08}.}
\begin{ruledtabular}
\begin{tabular}{c||rrrrr|ccc}
($\alpha$,$\beta$) & $[1,0]$ & $[1,1]$ & $[2,0]$ & $[2,1]$ & $[2,2]$ &
    {\bf I} & $\sigma_d$ & $\sigma_y$ \\ \hline
(1,1) &54.1     &39.1    &$-$89.2 &16.2    &3.3     &+   &+   & +(2,2) \\
(2,2) &41.6     &39.1    &38.9    &$-$8.4  &3.3     &+   &+   & +(1,1) \\
(3,3) &$-$109.7 &337.9   &14.3    &$-$16.4 &$-$6.1  &+   &+   & + \\
(1,2) &0        &122.5   &0       &22.9    &$-$0.5  &+   &$-$ & + \\
(1,3) &$-$346.3 &$-$26.1 &10.5    &$-$4.8  &$-$11.6 &$-$ &$+$ & $-$(2,3) \\
(2,3) &0        &26.1    &0       &22.4    &11.6    &$-$ &$-$ & $-$(1,3) \\
\end{tabular}
\end{ruledtabular}
\end{table}

The multi-orbital local correlations are described by
\begin{align}
\hH_I&=U\sum_{i,\alpha}\hn_{i\alpha\upa}\hn_{i\alpha\dna}
    +\left(U'-{1\over 2}J\right)\sum_{i,\alpha<\beta}\hn_{i\alpha}\hn_{i\beta}
    \label{h1} \\
&-J\sum_{i,\alpha\neq\beta}{\bf S}_{i\alpha}\cdot {\bf S}_{i\beta}
    +J\sum_{i,\alpha\neq\beta}c^\dg_{i\alpha\upa}
    c^\dg_{i\alpha\dna}c_{i\beta\dna}c_{i\beta\upa},
\nonumber
\end{align}
with $U=U'+2J$. In the perturbative treatment, the interactions are
decoupled in the HF approximation,
\[
\la c^\dg_{i\as} c_{i\beta\sigma'} \ra ={1\over 2}[n_\alpha+\sigma m_\alpha
\cos{(\bQ\cdot{\bf r}_i)}]\delta_{\alpha\beta}\delta_{\sigma\sigma'},
\]
where $n_\alpha$ and $m_\alpha$ the charge density and magnetization
on orbital $\alpha$. They are determined self-consistently in the
internal orbital-dependent fields in the charge and spin sector
respectively: $\Delta_\alpha={1\over 2}(2U-5J)n -{1\over 2}(U-5J)
n_\alpha$, $h_\alpha={1\over 2}Jm+{1\over 2}(U-J) m_\alpha$.
Clearly, $\Delta_\alpha$ renormalizes the crystal field splitting
and governs the orbital carrier transfer \cite{zhou05} while
$h_\alpha$ controls the distribution of the magnetic moment over the
three orbitals. The HF phase diagram is shown in Fig. 2a on the
$U$-$J$ plane. There are two phase boundaries, $U_\text{SDW}$ and
$U_\text{MIT}$, separating the paramagnetic (PM) phase, the
$\bQ$-SDW metal, and the $\bQ$-SDW insulator, respectively. The
evolutions of the magnetization $m=\sum_\alpha m_\alpha$, the
density of states (DOS) at the Fermi level $N_F$, and the
ferro-orbital (FO) order $m_\text{FO}=n_{yz}-n_{xz}$ that lifts the
degeneracy between $d_{xz}$ and $d_{yz}$\cite{kubo09} are shown in
Fig.~2b as a function of the correlation strength along the
trajectory $U=5J$. Due to the absence of FS nesting, the PM state is
stable until $U_\text{SDW}\sim$1.1 eV where SDW and FO order develop
simultaneously, followed by a MIT at $U_\text{MIT}\sim$1.9 eV beyond
which all FS pockets are fully gapped.
The origin of the insulating SDW phase is worth noting. The SDW
order is usually tied to the gapping of the band crossings (nodes)
upon folding by the ordering vector $\bQ$. However, the symmetry and
topology of the orbitals in the FeAs plane require that the nodes be
annihilated in sets of four \cite{ran09}. The six nodes in the LDA
band structure of LaOFeAs near $E_F$ implies a gapless SDW state
\cite{ran09}. We find that additional nodes are created in pairs as
the magnetization increases. The subsequent nodal annihilation in
sets of four produces a fully gapped SDW state. Indeed, for
$U=3\sim4$eV and $J=0.3\sim 0.8$eV appropriate for the Fe-pnictides,
the perturbative HF theory would predict an insulating SDW state
with large magnetization (also found in \cite{yu09}), contradicting
experimental observations. This is rooted in the fact that the HF
self-energies, i.e. the internal fields $\Delta_\alpha$ and
$h_\alpha$ scale with the correlation strengths due to
multioccupation and fail to capture correctly the correlation
effects.

\begin{figure}
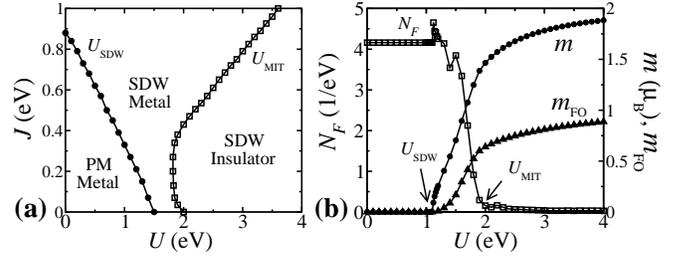

\begin{center}
\fig{3.4in}{fig2.eps} \vskip-2.4mm \caption{{\it HF results}. (a)
Phase diagrams on the $U$-$J$ plane. (b) Magnetization $m$, FO order
$m_\text{FO}$, and the Fermi level DOS $N_F$ as a function of $U$
along the trajectory $U=5J$.} \label{fig2}
\end{center}\vskip-8mm
\end{figure}

To treat interactions appropriately, it is necessary to take into
account that the probabilities for multi-occupation of the atomic
orbitals are energetically costly and significantly reduced even for
intermediate correlations. Since the $\bQ$-SDW is collinear, it is
sufficient to focus on the spin-dependent density-density
interactions in Eq.~(\ref{h1}). The atomic multiplets are thus the
$4^M$ ($M=3$ is the number of orbitals) local Fock states
$|\Gamma\ra= \prod_{\as} \left(c^\dg_{\as} \right)^{n^\Gamma_{\as}}
|0\ra$, where $n^\Gamma_{\as} =\la \Gamma |\hn_{\as}| \Gamma\ra=$0
or 1. The ground state can be obtained by Gutzwiller projected wave
function $|\Psi_G\ra=\hP_G |\Psi_0\ra$, where $|\Psi_0\ra$ is a
Slater determinate state and $\hP_G$ is the projection operator that
reduces the probability of multi-occupancy states in $|\Gamma\ra$.
The multiorbital Gutzwiller projection \cite{gebhard98} can be
formulated in the grand canonical ensemble,
\begin{equation}
\hP_G=\prod_i\hP_i,\quad \hP_i=\sum_\Gamma \left(\prod_{\as}
y^{n^\Gamma_{\as}}_{i\as} \right)\eta_{i\Gamma}\hat{m}_{i\Gamma},
\label{pg}
\end{equation}
where $\hat{m}_{i\Gamma} = |i\Gamma\ra \la i\Gamma| = \prod_{\as}
\left( \hn_{i\as}\right)^{n^\Gamma_\as}
\left(1-\hn_{i\as}\right)^{1-n^\Gamma_\as}$ projects onto the local
Fock state $|i\Gamma\ra$ and $\eta_{i\Gamma}$ is the probability
weighting factor determined variationally.
The density operator $\hn_{i\as}= \sum_{\Gamma} n^\Gamma_\as
\hat{m}_{i\Gamma}$. In Eq.~(\ref{pg}), the spin-orbital dependent
local fugacities $y_{i\as}$ maintain the charge density under the
projection, i.e., $n^0_{i\as}
=n_{i\as}=\sum_{\Gamma} n^\Gamma_\as m_{i\Gamma}$, with
$m_{i\Gamma}=\la\Psi_G| \hat{m}_{i\Gamma} |\Psi_G\ra$. The
projection is conveniently implemented using the Gutzwiller
approximation (GA) \cite{gebhard98} and is taken into account
locally by the statistical weighting factor multiplying the quantum
coherent state. For the hopping term, we find $\la\Psi_G|
c^\dg_{i\as} c_{j\beta\sigma} |\Psi_G\ra = g^\sigma_{i\alpha,j\beta}
\la\Psi_0| c^\dg_{i\as} c_{j\beta\sigma} |\Psi_0\ra$, where the
Gutzwiller factor $g^\sigma_{i\alpha,j\beta}=g_{i\as}
g_{j\beta\sigma}$, $g_{i\as}=\la\Psi_0| \hat{W}_{i\as} |\Psi_0\ra/
\la\Psi_0| \hP^2_i |\Psi_0\ra$ with $\hat{W}_{i\as}c^\dg_{i\as} =
\hP_i c^\dg_{i\as} \hP_i$,
\begin{equation}
g_{i\as}={1\over\sqrt{n^0_{i\as}\left(1-n^0_{i\as}\right)}}
\sum_{\Gamma,\Gamma'} D^\as_{\Gamma\Gamma'} \sqrt{m_{i\Gamma}
m_{i\Gamma'}}. \label{gfactor}
\end{equation}
Here $D^\as_{\Gamma\Gamma'} = \la\Gamma| c^\dg_\as |\Gamma'\ra=$0 or
1 describes the entanglement between the two multiplets. Hence the
projection leads to a nonperturbatively renormalized Hamiltonian,
\begin{equation}
\hH_\text{GA}= \sum_\bk
K_{s'\beta,\bk}^{s\alpha,\sigma}c^\dg_{s\alpha,\bk\sigma}
c_{s'\beta,\bk\sigma} + \tilde{E}_{s\Gamma} m_{s\Gamma} \label{hga}
\end{equation}
where $s$ labels the two-sublattice in the \bQ-SDW state, repeated
indices are summed, and $K^{s\alpha,\sigma}_{s'\beta,\bk} =
g^\sigma_{s\alpha,s^\prime\beta}\varepsilon_{s\alpha,s'\beta}
(\bk)+(\Delta^0_\alpha +\epsilon_{s\as}) \delta_{ss'}\delta_{\alpha\beta}$
with $\bk$ in the reduced BZ. The $\varepsilon_{s\alpha,s'\beta}(\bk)$ are
the LDA band dispersions of $\hH_0$. The main correlation effects are the
orbital dependent bandwidth reduction by the Gutzwiller factor; and a
renormalization of the crystal field splitting $\Delta_\alpha^0$ by
$\epsilon_{s\as}$ that originate from the fugacities and scale with the
kinetic energy. Note that, in contrast to the HF approximation, there are
no internal fields or self-energies proportional to the correlation
strengths for the fermionic quasiparticles. $U$ and $J$ only appear in the
energy level $E_\Gamma$ of the atomic multiplets $m_{s\Gamma}$ in
Eq.~(\ref{hga}) where $\tilde{E}_{s\Gamma} =E_\Gamma
-\sum_{\as}\epsilon_{s\as} n^\Gamma_{\as}$. The variational parameters
$\{m_{s\Gamma}, \epsilon_{s\as}\}$ are determined self-consistently by
minimizing the ground state energy of $\hH_{GA}$ under the completeness
condition $\sum_\Gamma \hat{m}_{i\Gamma}=1$.

\begin{figure}
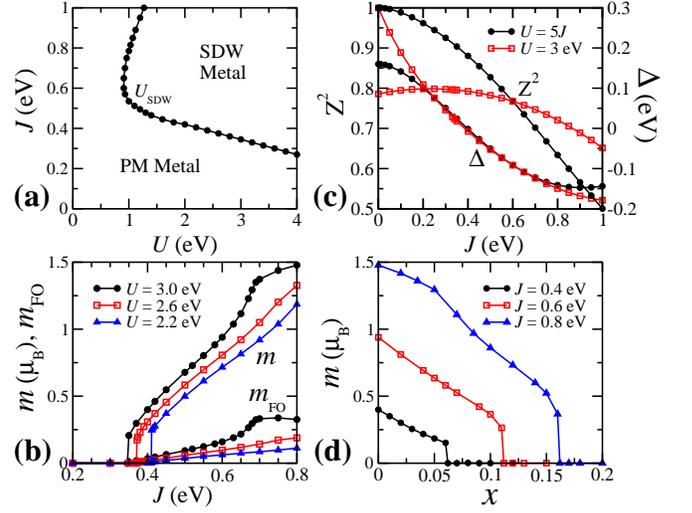

\begin{center}
\fig{3.4in}{fig3.eps} \vskip-2.4mm \caption{(color online). {\it GA
results}. (a) Phase diagram. (b) Magnetization $m$ and FO order
$m_\text{FO}$. (c) band-narrowing factor $Z^2$ and renormalized
crystal field splitting $\Delta$ as a function of $J$ at $U=3$eV and
for $U=5J$. (a)-(c) are for undoped case. (d) Electron doping $x$
dependence of $m$ at $U=3$eV.} \label{fig3}
\end{center}
\vskip-8mm
\end{figure}

The nonperturbative phase diagram is shown in Fig.~3a where a metallic
$\bQ$-SDW phase emerges at intermediate values of $(U,J)$ appropriate for
the pnictides. The gaplessness or the metallicity of this phase is
stabilized by the correlation effects taken into account nonperturbatively in the Gutzwiller approach. The crystal field
renormalization in Eq.~(\ref{hga}) counteracts the band splitting due to
magnetization. The insulating SDW phase erroneously appeared in the HF
theory (Fig.~2a) has a much higher energy and is absent in this regime. The
phase boundary has an interesting wedged shape indicating that the
transition is predominantly $J$ driven for intermediate $U$ and $U$ driven
for large $J$. Fig.~3b shows that the ordered moment
depends sensitively on the Hund's rule coupling $J$ and varies from
$0.3\mu_B$ to $1.5\mu_B$ in the range $J=0.3\sim0.8$eV for a range
of intermediate $U$. Surprisingly, the ordered moment depends weakly
on $U$ as can be seen from Fig.~3b. This indicates that in the
intermediate regime of correlations, the ordered moment is
controlled by the increased overlap with the atomic state of higher
spins due to the Hund's rule coupling instead of the localization of
the carriers due to Coulomb-blocking $U$. Note that the orbital order
is quite large in the 3-band model shown in Fig.~3b. For $U=3$eV, the relative orbital polarization $2m_\text{FO}/(n_{xz}+n_{yz})$ ranges from near 8\% at $J=0.5$eV to over 20\% near $J=0.8$eV. It has been shown recently \cite{chenetal} that such significant orbital ordering can partially account for the orthorhombic anisotropy of the reconstructed FS observed by ARPES \cite{shimojima}.

The different roles played
by $U$ and $J$ can also be seen in the PM phase. In Fig.~3c, we plot
the average band-narrowing factor $Z^2={1\over M}\sum_\alpha
g^\sigma_\alpha$ \cite{zfang09} and the renormalized crystal field
splitting $\Delta= \Delta_0+ \epsilon_{xy}-\epsilon_{xz,yz}$ as a
function of $J$ for both $U=3$eV and $U=5J$. The correlation induced
bandwidth reduction is predominantly controlled by $U$ and only
weakly depends on $J$. The crystal field renormalization is, on the
other hand, controlled by the Hund's rule coupling $J$ and is
essentially independent of $U$ in favor of multiorbital occupation
in a wide range of intermediate correlation strengths. Since the
intermediate correlation strengths of the pnictides are close to the
phase boundary shown in Fig.~3a, applied pressure would increase the
wave function overlap and reduce the correlation effects, which may
drive the system out of the SDW into the PM phase as observed
experimentally \cite{pressure}.

\begin{figure}
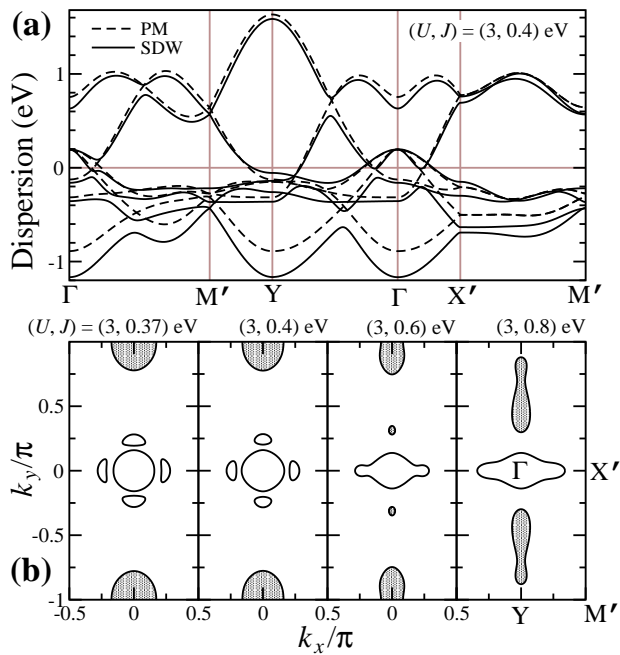

\begin{center}
\fig{3.2in}{fig4.eps} \vskip-2.4mm \caption{(a) Band dispersions of
the PM and SDW states at $U=3$ eV and $J=0.4$ eV. (b) FS at four
sets of interaction parameters with, from left to right, $m=$ 0.3,
0.4, 0.94, and 1.48$\mu_B$ respectively. Filled FS pockets are
electron-like and open ones are hole-like.} \label{fig4}
\end{center}
\vskip-8mm
\end{figure}

The band dispersions of the PM and SDW states at $U=3$ eV and
$J=0.4$ eV are shown in Fig.~4a. The overall bandwidth is somewhat
larger in the SDW phase due to the band splitting by magnetic order.
Fig.~4b shows the FS in the SDW state with increasing Hund's rule
coupling $J$ and the corresponding ordered moment. The FS topology
are rather sensitive to the magnetization. When the SDW moment is
small, the inner hole pocket around $\Gamma$ and the electron pocket
around Y remains almost unaltered, while the outer hole pocket and
the electron pocket around X suffer strong scattering via $\bQ$, and
become mostly gapped out, leaving behind four small hole pockets
around $\Gamma$ reflecting the Dirac cone-like band crossings. The
sizes of the small hole pockets, totaling $\sim2$\% of the reduced
BZ, are in reasonable agreement with quantum oscillation experiments
on Sr122 and Ba122 \cite{quan-oscil}. With increasing SDW moment,
the two small hole pockets along $\Gamma$-X move toward the central
hole pocket and eventually disappear as the Dirac cones are
annihilated. The hole pockets along $\Gamma$-Y are instead pushed
toward the electron FS near Y, changing character to being
electron-like through two Dirac nodes, and eventually coalesce to
form one set of elongated electron pockets as shown in Fig.~4b. As a
result, the FS topology measured by ARPES \cite{arpesSDW} would
depend rather sensitively on the magnetization in the surface
layers.

In summary, we have shown that the multiorbital Hubbard model in the
intermediate correlation regime correctly captures the important
correlation effects on the electronic structure and the SDW order in the
iron-pnictides. A nonperturbative treatment of the interactions is
essential to describe this correlated metallic state and the intricate,
complementary roles of the Coulomb repulsion and the Hund's rule coupling
generic to multiorbital systems. The calculated
magnetization is shown in Fig.~3d as a function of the electron doping concentration $x$ at $U=3$eV for several values of $J$. Despite the increase of the spin
susceptibility upon doping for our tight binding parameters, the
magnetization $m$ decreases monotonically with electron doping.
The results for the case of $U=3$eV and $J=0.4$eV describes well
the destruction of SDW order observed experimentally in LaO$_{1-x}$F$_x$FeAs.

We thank Chunhua Li for useful discussions. This work is supported
in part by DOE DE-SC0002554, NSF DMR-0704545, and the State of
Florida.

\end{document}